\documentclass[journal]{article}                                  

\usepackage{arxiv}
\usepackage{hyperref}
\usepackage{graphicx}          
\usepackage{amsmath} 
\usepackage{amssymb}  

\usepackage{amsthm}
\usepackage{color,soul}
\usepackage{flushend}
\usepackage{cite}
\usepackage{mfirstuc}
\usepackage{mathtools}
\usepackage{xcolor}
\usepackage{array,multirow}
\usepackage{algorithm} 
\usepackage{algorithmic}  
\usepackage[linesnumbered,algo2e,ruled,vlined,norelsize]{algorithm2e} 

\allowdisplaybreaks
\usepackage{tikz}
\usepackage{textcomp}
\usepackage{hyperref}
\usepackage{lipsum}

\newcommand\copyrighttext{%
  \footnotesize \textcopyright  This work is under review.
  Permission from the authors must be obtained for all other uses, in any current or future 
  media, including reprinting/republishing this material for advertising or promotional 
  purposes, creating new collective works, for resale or redistribution to servers or 
  lists, or reuse of any copyrighted component of this work in other works. 
  DOI: \href{<http://tex.stackexchange.com>}{To be generated.}}
\newcommand\copyrightnotice{%
\begin{tikzpicture}[remember picture,overlay]
\node[anchor=south,yshift=10pt] at (current page.south) {\fbox{\parbox{\dimexpr\textwidth-\fboxsep-\fboxrule\relax}{\copyrighttext}}};
\end{tikzpicture}%
}

\usepackage{authblk}
\title{\LARGE \bf
Transfer Learning Assisted XgBoost For Adaptable Cyberattack Detection In Battery Packs 
}

\author[1]{Sanchita Ghosh}
\author[1]{Tanushree Roy}
\affil[1]{Department of  Mechanical Engineering, Texas Tech University, Lubbock, TX 79409, US. Emails:~{\tt\small sancghos@ttu.edu, tanushree.roy@ttu.edu}.}

\begin{document}

\copyrightnotice

\maketitle
\pagestyle{empty}


\begin{abstract}
 Optimal charging of electric vehicle (EVs) depends heavily on reliable sensor measurements from the battery pack to the cloud-controller of the smart charging station. However, an adversary could corrupt the voltage sensor data during transmission, potentially causing local to wide-scale disruptions.  
 Therefore, it is essential to detect sensor cyberattacks in real-time to ensure secure EV charging, and the developed algorithms must be readily adaptable to variations, including pack configurations.
 To tackle these challenges, we propose adaptable fine-tuning of an XgBoost-based cell-level model using limited pack-level data to use for voltage prediction and residual generation.
  We used  battery cell and pack data from high-fidelity charging experiments in PyBaMM and `liionpack' package to train and test the detection algorithm. The  algorithm's performance has been evaluated for two large-format battery packs under sensor swapping and replay attacks. The simulation results also highlight the adaptability and efficacy of our proposed detection algorithm.
\end{abstract}


\section{Introduction}
As demand for emission-free transportation increases, smart charging and vehicle-to-grid technologies for EV will help balance loads, reduce grid stress, and support the integration of renewable energy sources. 
 However, the smart charging infrastructures are susceptible to cyberattacks causing unsafe battery operations in EV, financial/energy losses, and inducing voltage oscillations that can ultimately reduce grid resilience and cause power outages \cite{johnson2022cybersecurity,ghosh2024koopman,ghosh5028845detection}. In particular, corruption of sensor data (e.g., voltage, temperature) can adversely affect the charging control actions leading to over-/under- charged EV, faster battery degradation, or unstable grid operation \cite{carlson2021consequence}. Thus, it is crucial to detect the presence of the sensor attacks during charging at the earliest to ensure safe and reliable EV charging operations as well as resilient grid operations \cite{trevizan2022cyberphysical}.

While model-based methodologies have been explored for sensor attack detection, they require reliable battery knowledge and often fail to accommodate the changes in model parameters due to battery aging or different cell-chemistry. Alternatively, data-driven detection algorithms based on machine learning techniques do not rely on the system model but rather require pre-training with experimental battery data  (\cite{trevizan2022cyberphysical}). For instance, neural networks and long short-term memory (LSTM) networks have been used for state-of-charge estimation for detecting sensor cyberattacks on EV batteries \cite{rahman2018study,rao2023detection}. Additionally,  \cite{mitikiri2025anomaly} proposed an LSTM-based autencoder model to generate battery current prediction and utilized the Kolmogorov–Smirnov statistic distance between the predicted and observed data to detect cyberattacks on EV batteries.  Moreover, \cite{basnet2020deep} conducted a performance comparison between deep-neural-network and LSTM-based intrusion detection models for EV charging attacks.  \cite{haider2020data} also adopted a time-series clustering approach to monitor battery operation data and detect the EV charging attack. 

Nevertheless, such machine learning models highly rely on both the volume and the quality of available data while real battery charging data is often kept confidential \cite{ronanki2023electric}. Specifically for battery packs, generating experimental battery data is expensive, fatiguing, time-consuming, and requires special hardware set-up \cite{channegowda2022synthetic}. 
In addition, machine learning based models fail to adapt and generalize across various operating conditions or pack configurations due to the distribution discrepancy between the training and testing data \cite{shen2023transfer}. While the Koopman operator-based online learning algorithm proposed by \cite{ghosh2024koopman} offers a generalizable and adaptable approach for detecting cyberattacks in battery systems, it is computationally expensive and not well-suited for scaling to larger battery systems. To address these challenges, lately, researchers have adopted transfer learning (TL) to ensure reliable cross-domain battery state estimation \cite{fu2024data}. In particular, researchers utilized TL for state-of-charge (\cite{qin2021transfer}), state-of-health (\cite{tan2019transfer}), and capacity (\cite{fu2024data})  estimation (vide references in \cite{shen2023transfer, liu2023transfer}). However, to the best of our knowledge, this powerful TL tool has not yet been adopted for cybersecurity analysis for EV battery charging. Furthermore, these papers focused on employing TL with deep/convolutional neural networks and LSTM networks which are computationally expensive and prone to over-/under-fitting with TL \cite{liu2023transfer}. 
Moreover, with the battery pack data insufficiency, there is a research gap to leverage TL for effectively transferring intrinsic battery charging dynamics from the cell-level data to the pack-level and use to for cyberattack detection in large-format EV battery systems \cite{hosen2022strategic}. 
To address these research gaps, the main contributions of this paper are as follows:
\begin{enumerate}
    \item We proposed a TL-assisted XgBoost model generation using baseline battery cell-level data with limited pack data, to predict battery pack-level voltage. The prediction from this TL-XgBoost (TL-XGB) model is used for residual generation to detect EV charging cyberattacks at charging stations.  
    \item The proposed algorithm is computationally efficient, easily adaptable to different pack configurations, and capable of real-time online detection. 
    \item We evaluated the algorithm performance for two different pack configurations while considering high-fidelity battery charging data generated with PyBaMM and liionpack, under realistic sampling frequency and measurement noise. 
\end{enumerate}

The rest of the paper is organized as follows: Section~\ref{prob_from} explains the framework for this work and Section~\ref{di_scheme} introduces the proposed TL-assisted XgBoost detection algorithm. Next,  we present the results of the algorithm's performance evaluations in Section~\ref{sim_rel}. Finally, Section~\ref{conc} concludes our work.

\section{Problem Framework} \label{prob_from}
\begin{figure}[h]
    \centering
    \includegraphics[width=0.7\linewidth]{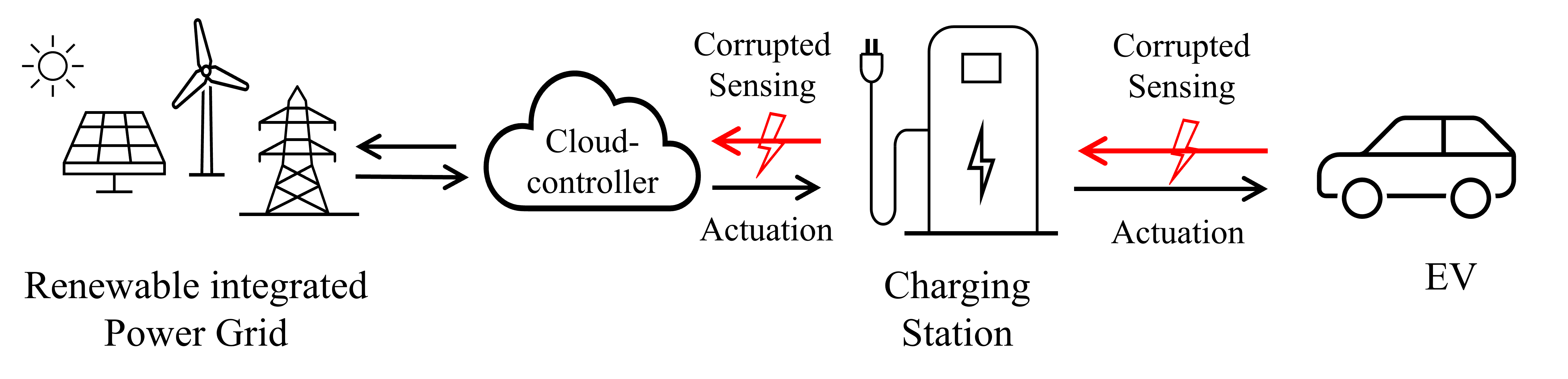}
    \caption{Overview of our problem framework.}
    \label{fig:oveeview}
\end{figure}

 A cloud-based EV charging controller monitors the terminal voltage measurements from the EV battery system and generates the appropriate charging current actuation signal to achieve optimal and safe charging of EVs while maintaining the grid stability \cite{erdogan2022multi}. This work addresses the problem of sensor attack detection while considering the vulnerabilities of such cloud-based EV charging at smart charging stations.
In this framework, an adversary can corrupt the voltage sensor measurements from the EV battery, as they are transmitted through communication networks \cite{johnson2022cybersecurity,ghosh2024koopman}. Rapid detection of such sensor attacks is crucial in this scenario, as inaccurate voltage measurements can lead to poor battery management or even unstable grid operations \cite{ghosh2024isolating}. Fig~\ref{fig:oveeview}  illustrates our problem framework and the potential attack vectors. 

 In the presence of sensor attack $\delta_{V_t}$, the  battery charging dynamics is defined as:
\begin{align} \label{vt_dynamics}
      & z (k+1) = g(z(k),I_c(k)), \\ & V_{t,m}(k)  = h(z(k),I_c(k)) + \delta_{V_t},
\end{align}
where $z (k) \in \mathbb{R}^d$ is the battery state vector which may contain the internal battery states such as lithium-ion concentrations in the solid and liquid phase, solid and liquid phase potentials, state-of-charge, open circuit potential, and temperature \cite{li2013reduced}. Then, $I_c (k) \in \mathbb{R}$ is the charging current input generated by the cloud-controller, and constant-current-constant-voltage (CCCV) charging policy that is frequently adopted in real-world operations to ensure improved battery health \cite{li2020optimized}.  $V_{t,m}(k) \in \mathbb{R}^q$ is the terminal voltage measurement at $k^{th}$ instant for $m^{th}$ module, $\forall m \in [1, \cdots, q]$. This module number $q$ can vary with pack configurations. For a battery cell, we can interpret $q = 1$ and we can rename $V_{t,m}(k)$ as $V_t (k)$ to denote the cell terminal voltage measurement. $h : \mathbb{R}^d \rightarrow \mathbb{R}^q$ denotes the nonlinear output function.  Lastly, the vector field $g: \mathbb{R}^d \times \mathbb{} \rightarrow \mathbb{R}^d$  captures the internal dynamics of the battery.

In this work, we propose a TL-assisted XgBoost detection algorithm to detect the presence of sensor attack $\delta_{V_t}$. In particular, we first utilize the battery cell-level data to train a base XgBoost model. Next, we adopt TL technique to fine-tune the base model with limited pack-level data, to ensure adaptability across various configurations of large-format battery packs. The TL-XGB model is then used to generate module voltage prediction from time-series voltage and current data, i.\,e., the model utilizes $V_{t,m} (k)$ and $I_c (k)$ data as input to generate a voltage prediction for $V_{t,m} (k+1)$. This prediction is sent to a residual generator that calculates the deviation in generated prediction from the module voltage measurements to detect the presence of sensor attacks. 
In addition, due to the residual-based detection strategy, the algorithm does not require pre-training with attacked data and thus, reliably performs against unforeseen cyberattacks. The proposed algorithm is validated with two different battery pack configurations against presently prevalent EV sensor charging cyberattacks such as false-data-injection (FDI) and replay attacks.

\section{Cyberattack Detection Algorithm} \label{di_scheme}
\begin{figure*}[t]
    \centering
    \includegraphics[width=0.7\linewidth]{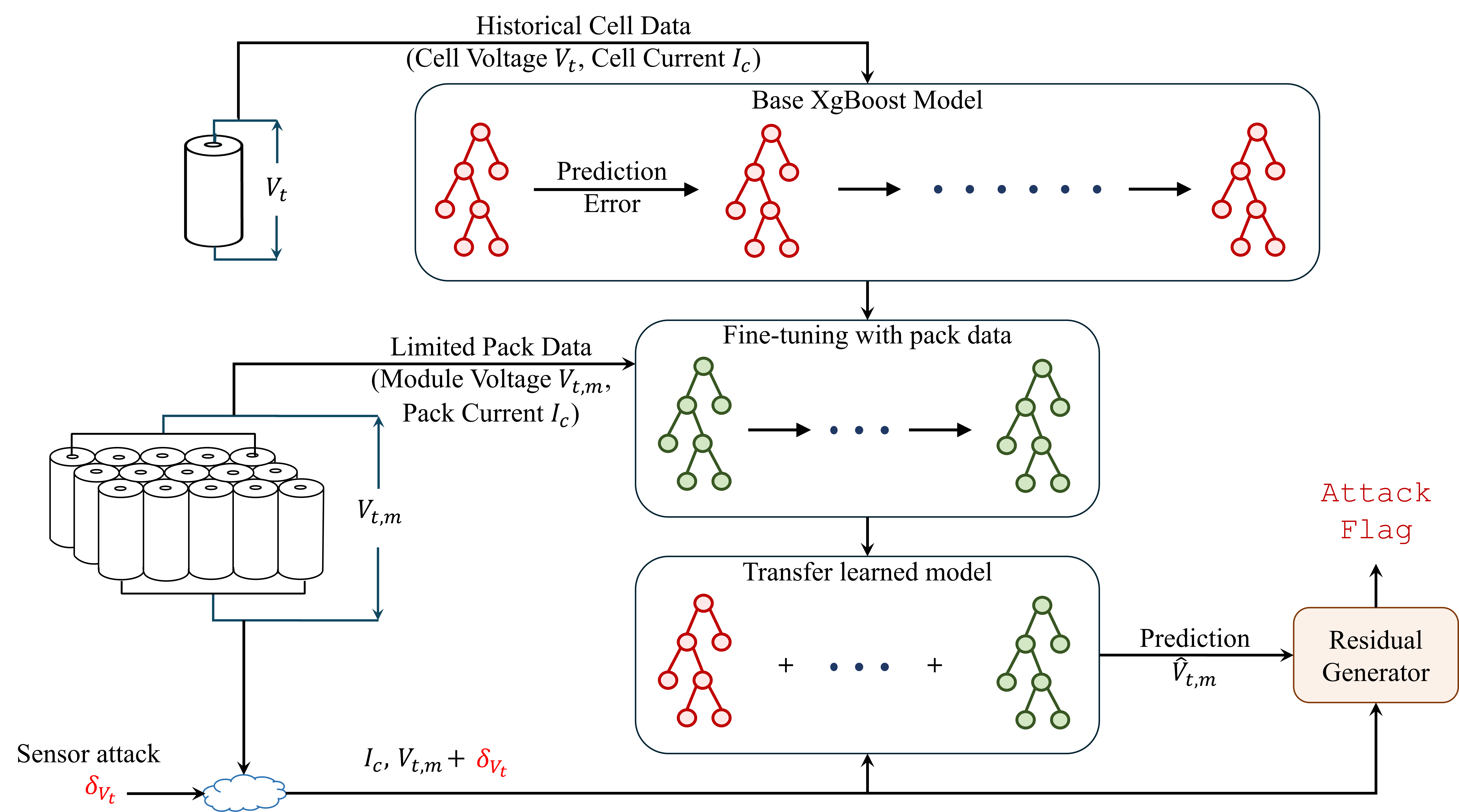}
    \caption{Block diagram illustrates the proposed cyberattack detection framework with transfer learning assisted XgBoost.}
    \label{fig:block}
\end{figure*}

The first step of implementing the algorithm is to obtain the experimental dataset for the XgBoost base model (from cell) and the TL-XGB model (from packs). Next, the base and TL-XGB models are trained with this data. These models are used to develop our detection algorithm in the third subsection. Fig~\ref{fig:block} shows the block diagram illustrating the proposed TL-assisted XgBoost detection algorithm for sensor attacks on EV charging.
\subsection{Data generation with PyBaMM and liionpack}\label{dataGen}
We have utilized the open-source battery simulation package Python Battery Mathematical Modeling (PyBaMM) [\cite{sulzer2021python}] and `liionpack' [\cite{tranter2022liionpack}] to generate our battery charging scenarios respectively for cells and large-format packs. 
PyBaMM and liionpack have become reliable tools for battery cybersecurity research to evade unsafe charging operations with costly experimental setup for repeated testing  of batteries under cyberattack  \cite{jin2024assessing}.
Table~\ref{tab:pybammP}  lists the configurations considered during our data generation using PyBaMM and liionpack. For both battery cell and packs, we have considered the commercial cylindrical battery cell LGM50 with NMC 811 ($80\%$ Nickel, $10\%$ Manganese, and $10\%$ Cobalt) as positive electrode and bi-component Graphite-$SiO_x$ as negative electrode \cite{chen2020development}. In PyBaMM and liionpack, we used the Single-Particle Model (SPM) during our battery charging data generation \cite{atlung1979dynamic}. The SPM model can capture the internal electrochemical mechanics of the battery cell and renders good accuracy with computationally efficient operation. In addition, we adopt the traditional CCCV  charging policy for our charging data generation. With this setup, we generate a large and enriched baseline cell-level dataset containing charging data for different operating conditions, e.g.,  charging span, current rate, initial state-of-charge, and temperature. We generate two brief-span (15 minutes) charging cycle data for each pack with 0.8 and 1.2 C-rates for training of our TL-XGB models, while we generate testing data with 1C rate charging for both packs to conduct our sensor attack case studies in Section~\ref{sim_rel}. We explain the training of our base and TL-XGB models in the next section.

\begin{table}[h!]
    \centering
    \begin{tabular}{|c|c|c|c|}
    \hline
       Parameters  & Cell & Pack 1 & Pack 2 \\
       \hline
        Pack  configuration & $-$ & 20p100s & 25p80s  \\
        \hline
        Maximum voltage & 4.2 $V$ & 424 $V$ & 338 $V$ \\
        \hline
        Discharge capacity & 5 $Ah$ & $100Ah$ & 125 $Ah$\\
        \hline
         Power capacity & 18 $Wh$ & 35 $kWh$ & 35 $kWh$ \\
         \hline
         Input current data & Cell  & Pack & Pack \\
         \hline
         Voltage sensor data & Cell & 4  modules & 5  modules \\
         \hline
         Measurement noise &  \multicolumn{3}{c|}{$\mathcal{N}(0,\sigma^2)$,\,\,\, $\sigma = \pm 0.1\% \, V$} \\
         \hline
         Sampling rate &  \multicolumn{3}{c|}{1 $Hz$}\\
         \hline
    \end{tabular}
    \caption{Battery charging data  specifications}
    \label{tab:pybammP}
    \vspace{-3mm}
\end{table}
\subsection{Base \& TL-XGB model generation}
XgBoost (eXtreme Gradient Boosting) is a tree-based supervised machine learning algorithm that sequentially trains an ensemble of trees, and each tree training assimilates the error from the previous tree predictions. The model output at training iteration step $l$ is defined as follows:
\begin{align}
    \hat{y}_i^l = \sum\limits_{j=1}^l f_j(x_i) = \hat{y}_i^{l-1} + f_l(x_i),
\end{align}
where, $x_i$ denotes the input features of the $i^{th}$ training data point, $\hat{y}_i^l$ is the corresponding model output at $l^{th}$ iteration step and $i\in [1,\cdots,N]$ where $N$ is the total number of training data points.  The function $f_j$ contains the structure and influence of the $j^{th}$ tree. Hence,  the model incorporates one additional tree $f_l$ at each iteration step $l$. We have considered the squared error loss function
\begin{align}
    \mathcal{L}(y, \hat{y}) = \sum\limits_i \left(y_i - \hat{y}_i \right)^2,
\end{align}
for training the XgBoost algorithm.
$y_i$ is the target value corresponding to $x_i$ data point. With this loss function $\mathcal{L}$, the objective function at iteration step $l$ is defined as:
\begin{align}
    J_{l} & = \sum\limits_i \mathcal{L} \left(y_i, \, \hat{y}_i^{l-1} + f_l(x_i) \right) + \Omega(f_l).
\end{align}
Here, the regularization term $\Omega(f_l)$ defines the complexity of $l^{th}$ tree. Now, the objective function can be re-written as \eqref{tobj} by using Taylor expansion and removing constant terms,\
\begin{align}\label{tobj}
    J_{l} & = \sum\limits_i \left[ m_i \, f_l(x_i) + \frac{1}{2} n_i f_l^2(x_i) \right] + \Omega(f_l),
\end{align}
where $m_i = \partial_{\hat{y}_i^{l-1}} \mathcal{L} \left(y_i, \, \hat{y}_i^{l-1} \right)$ and $ n_i = \partial_{\hat{y}_i^{l-1}}^2 \mathcal{L} \left(y_i, \, \hat{y}_i^{l-1} \right)$. 
Thus, XgBoost consistently boosts the tree-training process under regularization by rectifying the prediction error generated by the previous ensemble of trees. In addition, XgBoost provides a highly computationally efficient model, especially for large datasets.  For our framework, we first train a base XgBoost model with the generated baseline cell dataset. Furthermore, we utilize the voltage and current data at $k^{th}$ instant as input feature with next instant voltage measurement as the corresponding target, i.\,e., the input feature is given by $x_i = \left[V_t (k), \,\, I_c (k)\right]$  and the corresponding target is given by $y_i = V_t (k+1)$.  We use a training data of size $N = 81\times10^3$  for this base model training and a validation data set of size $N = 13\times10^3$. Moreover, the base model has 400 numbers of trees with a maximum depth of 4, i.e., each tree can divide up to 4 times. We also tested the base model performance where the model exhibits a maximum absolute error \% of 0.003.  Table~\ref{tab:xgboost} lists the hyper-parameters for this training.

\begin{figure}[h!]
    \centering
    \includegraphics[width=0.6\linewidth]{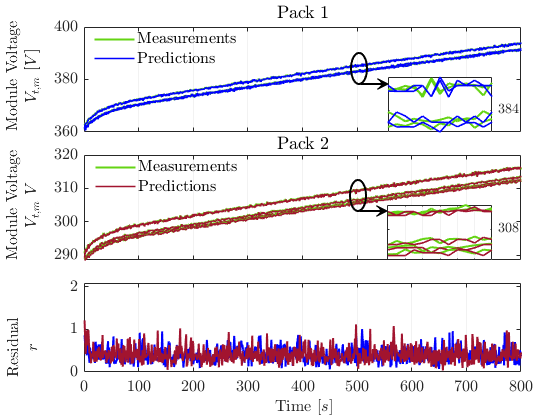}
    \caption{Under nominal EV charging operation, the plot exhibits measured and predicted module voltages for pack 1 (top) and pack 2 (middle) followed by the generated residual $r$ (bottom).}
    \label{fig:pred_acc}
\end{figure}

Next, we fine-tune the base model separately for two battery packs with the corresponding limited charging data to obtain their respective TL-XGB models. For each pack, we use one module data for testing, one for validation, and the rest of the module data 
as training data for both packs (i.\,e., $x_i = \left[V_{t,m} (k), \,\, I_c (k)\right]$ is the input feature and $y_i = V_{t,m} (k+1)$ is the target).  Battery pack~1 has a 20p100s configuration, i.\,e.  $q = 4$ parallel modules, where each module has 5 parallel branches with 100 series cells each. Notably, for pack~1, we use a limited training sample of size $N = 1800$  and a validation dataset of size 900. With this dataset, the base XgBoost model is fine-tuned with an additional 3 trees of a maximum depth of 2 to obtain the TL-XGB model for pack~1.  Under nominal charging, we show the true module voltage measurements and their voltage prediction from the TL-XGB model for pack~1 in the top plot of Fig~\ref{fig:pred_acc}. The proposed TL-XGB model provides reliable module voltage predictions as shown in the zoomed inset of this plot. During testing, the TL-XGB model exhibits a maximum absolute error \% of 0.004.  We next use this TL-XGB model for pack~1 to generate the module voltage prediction for attack detection. 

Pack~2 considered in our case study has the configuration of 25p80s,  i.\,e.  $q = 5$ parallel modules such that each module has 5 parallel branches with 80 series cells each.  For this pack, we use a training dataset of size $N = 2700$  and a validation dataset of size 900. To obtain the TL-XGB model for pack~2, the base model is fine-tuned with an additional 2 trees of maximum depth 8 using the training and validation dataset. Similarly, we use this second TL-XGB model to generate module voltage prediction for pack~2. The second plot of Fig~\ref{fig:pred_acc} shows the measured and predicted module voltages with TL-XGB model under nominal charging for this pack.  This TL-XGB model for pack 2 also provides highly accurate predictions with a maximum absolute error \% of 0.003 as shown in the zoomed inset of this plot.    Table~\ref{tab:xgboost} shows the hyper-parameters for the base XgBoost model and the two TL-XGB models corresponding to each pack. Comparing the hyper-parameters of the base model with the TL-XGB models, it is evident that the proposed transfer learning strategy greatly reduces the additional model complexity and data requirement. Moreover, the runtime for the pack-level TL adaption is approximately a quarter of a second, which shows the quick adaptability and computational efficiency of this method for different pack configurations.

\begin{table}[h!]
    \centering
    \begin{tabular}{|c|c|c|c|}
    \hline
         & Base  & TL: Pack 1 & TL: Pack 2 \\
       \hline
        No of trees & 400 & 3 & 2  \\
        \hline
        {Maximum depth} & 4  & 2  & 8  \\
        \hline
        Learning rate & 0.12 & 0.035 & 0.02\\
        \hline\hline
         Training   data size & 81000   & 1800  & 2700 \\
         \hline
        Validation  data size & 13000  & 900 & 900 \\
         \hline
         Runtime  $[s]$ & 5.12  & 0.21 & 0.29 \\
         \hline
         Maximum abs error \% & 0.003  & 0.004 & 0.003 \\
         \hline
    \end{tabular}
    \caption{Table showing  training hyper-parameters and  model performances metrics. }
    \label{tab:xgboost}
    \vspace{-3mm}
\end{table}

\subsection{Attack detection}
We utilize the TL-XGB models to generate online module voltage prediction $\widehat{V}_t(k)$ at $k^{th}$ instant. Borrowing techniques from diagnostics theory, let us define the residual $r$ as the maximum error in voltage prediction for all modules \cite{Ding}. Mathematically, this can be written as:
\begin{align}
    r(k) = \max\limits_{m} \lvert V_{t,m}(k) - \widehat{V}_{t,m}(k) \rvert ; \quad \forall m \in [1, \cdots, q].
\end{align}
 Appropriate threshold selection is crucial to ensure the real-time detection performance of the proposed algorithm. In particular, we intend to choose an optimum threshold value to minimize misdetection under cyberattacks while minimizing the false alarm rate due to measurement noises under nominal charging as well \cite{Ding}. For our proposed algorithm, under nominal EV charging operation,  the generated residual $r$  is non-zero but small due to the data uncertainty and the TL-XGB models' prediction error. The last plot of Fig~\ref{fig:pred_acc}  shows this error as the generated residuals, which have the maximum value of 1.5 for both packs under nominal scenarios.  Based on this maximum value, we define a threshold $\epsilon=2$ for both packs to account for some practical variability and flexibility. Next, we compare the residual $r$ with this threshold $\epsilon$ to generate an attack flag under sensor attack. 

Now, consider a scenario where an adversary injects a sensor attack from $k_{0}$ to $k_{f}$ instant, i.e., ${V}_{t,m} (k), \, k_0 \leqslant k \leqslant k_f $ is corrupted. However, the TL-XGB model generates the prediction $\widehat{V}_{t,m}(k_{0})$ based on the uncorrupted voltage measurement ${V}_{t,m}(k_{0}-1)$. Thus, the predicted $\widehat{V}_{t,m}(k_{0})$ does not match with the corrupted voltage ${V}_{t,m}(k_{0})$ and the residual $r$ no longer remains small. Consequently, residual $r (k_{0})$  crosses the threshold $\epsilon$. However, after that instant,  the model utilizes the corrupted voltage measurement to generate the corrupt prediction during $\delta_{V_t} \neq 0$. This implies that $r(k)$ stays below the threshold for $k_0 < k < k_f $. Once again, when the attack is withdrawn at $k_{f}$ instant, the TL-XGB model prediction $\widehat{V}_{t,m}(k_{f})$ is based on the corrupted measurement ${V}_{t,m}(k_{f}-1)$ which does not match with the uncorrupted measurement ${V}_{t,m} (k_{f}) $ and residual $r (k_{f})$  crosses the threshold $\epsilon$ again.    Thus, we first set the attack flag from 0 to 1 at $k = k_0$ when  $r (k_{0})$  crosses the threshold. When  $r (k_{f})$ crosses the threshold the next time, the attack flag is reset from 1 to 0 at $k =k_f$. This set and reset rule of our flag is given by:
\begin{align} \label{resd}
    & r(k) \geqslant \epsilon \, \Rightarrow \, \text{Attack Flag} = 1- \text{Attack Flag}. 
\end{align}
The algorithm detects a sensor attack for the time when attack flag is set to 1. 
 With this TL-assisted XgBoost detection algorithm, we present the results of our simulation case studies in the next section.

\section{Results and Discussions} \label{sim_rel}
In this section, we present performance evaluation for two case studies with presently prevalent EV charging attacks to demonstrate the efficacy of the proposed algorithm.

\subsection{Performance evaluation against data swap FDI attack}
\begin{figure}[h!]
    \centering
    \includegraphics[width=0.6\linewidth]{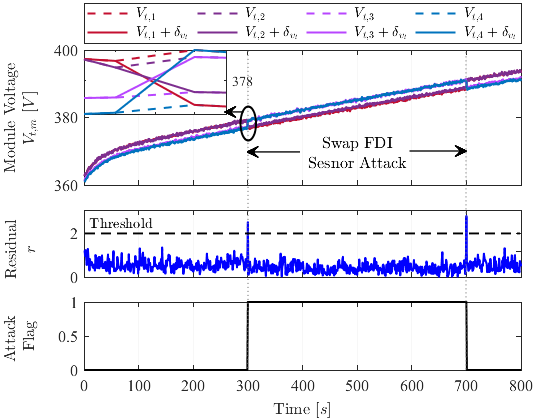}
    \caption{Under data swap FDI attack on pack~1, the plot exhibits nominal and corrupted module voltages for pack 1 (top), the generated residual $r$ (middle), and the attack flag (bottom).}
    \label{fig:dataSwap}
\end{figure}
In this scenario, we consider the pack~1 with 4 modules. A tailored FDI attack that swaps the  voltage data to rearrange the ascending module voltages to descending module voltages is injected. Such data swap attack hampers the cell-balancing functionality of the BMS, however
cell-balancing is crucial to ensure optimum battery performance with improved battery health \cite{khalid2021investigation}. The attack starts at $300$s and continues until $700$s. This phenomenon is captured in the first plot of Fig~\ref{fig:dataSwap} and the zoomed inset shows how the voltage of module 1 through 4 becomes the voltage of 4 through 1 respectively. The residual here is generated with \eqref{resd} where $\widehat{V}_{t,m}$ is obtained from the pack 1 TL-XGB model. The generated residual $r$ crosses the threshold during both attack initiation at $t = 300$ and withdrawal at $t =700$ as shown in the second plot of Fig~\ref{fig:dataSwap}. Consequently, the algorithm sets the attack flag to 1 for the attack duration to reliably detect the presence of the attack in the last plot of Fig~\ref{fig:dataSwap}.

\subsection{Performance evaluation against replay attack}
\begin{figure}[h!]
    \centering
    \includegraphics[width=0.6\linewidth]{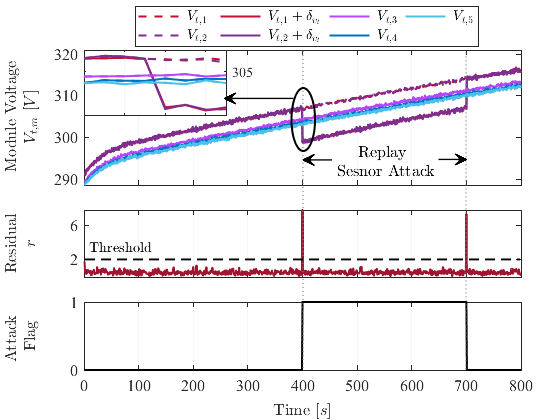}
    \caption{Under replay sensor attack on pack~2, the plot exhibits nominal and corrupted module voltages for pack 2 (top), the generated residual $r$ (middle), and the attack flag (bottom).}
    \label{fig:replay2p}
\end{figure}
Here, the second case study is performed on pack~2, we consider that the adversary first stores both the voltage sensor measurements starting from $100$s to $400$s to inject a replay attack by feeding back the recorded sequence from $400$s to $700$s. Moreover, we consider that the adversary only replays the recorded voltage measurement for module 1 and 2 as shown in the first plot of Fig~\ref{fig:replay2p}. The second plot of Fig~\ref{fig:replay2p} shows that the generated residual (calculated with \eqref{resd} and  pack 2 TL-XGB model) crosses the threshold at  $400^{\text{th}}$s and $700^{\text{th}}$s. Thus, the attack flag is set to 1 for the attack duration in the last plot of Fig~\ref{fig:replay2p}. The algorithm reliably detects this partial corruption of only two module voltage sensor data in this the replay attack scenario.

\section{Conclusion} \label{conc}
In this work, we propose an adaptable and pack configuration independent detection algorithm for sensor attack during EV charging by transfer learning cell-level XgBoost model for battery packs. Specifically, we trained the base XgBoost model with cell-level charging data and later fine-tuned it using TL with limited pack-level data, to transfer the intrinsic cell-level knowledge of the battery charging dynamics. We generated two TL-XGB models for the two pack configurations to generate predictions for their module voltages. These predictions are compared to the measured voltage to generate detection residual. Simulation results show that both sensor swapping attack and sensor replay attacks on the packs are reliably detected.


\bibstyle{arxiv}
\bibliography{ref.bib}

\begin{thebibliography}{10}

\bibitem{johnson2022cybersecurity}
Jay Johnson, Benjamin Anderson, Brian Wright, Jimmy Quiroz, Timothy Berg, Russell Graves, Josh Daley, Kandy Phan, Micheal Kunz, Rick Pratt, et~al.
\newblock \capitalisewords{Cybersecurity for Electric Vehicle Charging Infrastructure.}
\newblock Technical report, Sandia National Lab.(SNL-NM), Albuquerque, NM (United States), 2022.

\bibitem{ghosh2024koopman}
Sanchita Ghosh and Tanushree Roy.
\newblock \capitalisewords{Koopman Operator-based Detection-Isolation of Cyberattack: A Case Study on Electric Vehicle Charging}.
\newblock In {\em 2024 American Control Conference (ACC)}, pages 2236--2241. IEEE, 2024.

\bibitem{ghosh5028845detection}
Sanchita Ghosh and Tanushree Roy.
\newblock \capitalisewords{Detection and Isolation of Battery Charging Cyberattacks Via Koopman Operator}.
\newblock {\em Available at SSRN 5028845}, 2024.

\bibitem{carlson2021consequence}
R~Carlson.
\newblock \capitalisewords{Consequence-Driven Cybersecurity for High-Power EV Charging Infrastructure DOE Vehicle Technologies Program Annual Merit Review}.
\newblock {\em Proceedings of the DOE Vehicle Technologies Program Annual Merit Review, Washington, DC, USA}, 24, 2021.

\bibitem{trevizan2022cyberphysical}
Rodrigo~D Trevizan, James Obert, Valerio De~Angelis, Tu~A Nguyen, Vittal~S Rao, and Babu~R Chalamala.
\newblock \capitalisewords{Cyberphysical security of grid battery energy storage systems}.
\newblock {\em IEEE Access}, 10:59675--59722, 2022.

\bibitem{rahman2018study}
Syed Rahman, Haneen Aburub, Yemeserach Mekonnen, and Arif~I Sarwat.
\newblock \capitalisewords{A study of {EV BMS} cyber security based on neural network SOC prediction}.
\newblock In {\em 2018 IEEE/PES Transmission and Distribution Conference and Exposition (T\&D)}, pages 1--5. IEEE, 2018.

\bibitem{rao2023detection}
K~Dhananjay Rao, Manasa Taddi, Tharun Sriramula, Dilip~Kumar Baliga, Akhil Simhadri, and Parth~Sarathi Panigrahy.
\newblock \capitalisewords{Detection of Cyber Attacks on Wireless BMS of Electric Vehicles using Long Short-Term Memory Networks}.
\newblock In {\em 2023 7th International Conference on Computation System and Information Technology for Sustainable Solutions (CSITSS)}, pages 1--6. IEEE, 2023.

\bibitem{mitikiri2025anomaly}
Sagar~Babu Mitikiri, Vedantham~Lakshmi Srinivas, and Mayukha Pal.
\newblock \capitalisewords{Anomaly detection of adversarial cyber attacks on electric vehicle charging stations}.
\newblock {\em e-Prime-Advances in Electrical Engineering, Electronics and Energy}, 11:100911, 2025.

\bibitem{basnet2020deep}
Manoj Basnet and Mohd~Hasan Ali.
\newblock \capitalisewords{Deep learning-based intrusion detection system for electric vehicle charging station}.
\newblock In {\em 2020 2nd International Conference on Smart Power \& Internet Energy Systems (SPIES)}, pages 408--413. IEEE, 2020.

\bibitem{haider2020data}
Syed~Naeem Haider, Qianchuan Zhao, and Xueliang Li.
\newblock \capitalisewords{Data driven battery anomaly detection based on shape based clustering for the data centers class}.
\newblock {\em Journal of Energy Storage}, 29:101479, 2020.

\bibitem{ronanki2023electric}
Deepak Ronanki and Harish Karneddi.
\newblock \capitalisewords{Electric vehicle charging infrastructure: Review, cyber security considerations, potential impacts, countermeasures and future trends}.
\newblock {\em IEEE Journal of Emerging and Selected Topics in Power Electronics}, 2023.

\bibitem{channegowda2022synthetic}
Janamejaya Channegowda, Vinayak Raj~Urs, and Chaitanya Lingaraj.
\newblock \capitalisewords{Synthetic battery attribute generation to surmount data scarcity using auto-correlation mechanism}.
\newblock {\em International Journal of Energy Research}, 46(7):9882--9891, 2022.

\bibitem{shen2023transfer}
Liyuan Shen, Jingjing Li, Lichao Meng, Lei Zhu, and Heng~Tao Shen.
\newblock \capitalisewords{Transfer learning-based state of charge and state of health estimation for li-ion batteries: A review}.
\newblock {\em IEEE Transactions on Transportation Electrification}, 10(1):1465--1481, 2023.

\bibitem{fu2024data}
Shiyi Fu, Shengyu Tao, Hongtao Fan, Kun He, Xutao Liu, Yulin Tao, Junxiong Zuo, Xuan Zhang, Yu~Wang, and Yaojie Sun.
\newblock \capitalisewords{Data-driven capacity estimation for lithium-ion batteries with feature matching based transfer learning method}.
\newblock {\em Applied Energy}, 353:121991, 2024.

\bibitem{qin2021transfer}
Yan Qin, Stefan Adams, and Chau Yuen.
\newblock \capitalisewords{Transfer learning-based state of charge estimation for lithium-ion battery at varying ambient temperatures}.
\newblock {\em IEEE Transactions on Industrial Informatics}, 17(11):7304--7315, 2021.

\bibitem{tan2019transfer}
Yandan Tan and Guangcai Zhao.
\newblock \capitalisewords{Transfer learning with long short-term memory network for state-of-health prediction of lithium-ion batteries}.
\newblock {\em IEEE Transactions on Industrial Electronics}, 67(10):8723--8731, 2019.

\bibitem{liu2023transfer}
Kailong Liu, Qiao Peng, Yunhong Che, Yusheng Zheng, Kang Li, Remus Teodorescu, Dhammika Widanage, and Anup Barai.
\newblock \capitalisewords{Transfer learning for battery smarter state estimation and ageing prognostics: Recent progress, challenges, and prospects}.
\newblock {\em Advances in Applied Energy}, 9:100117, 2023.

\bibitem{hosen2022strategic}
Md~Sazzad Hosen, Ashkan Pirooz, Theodoros Kalogiannis, Jiacheng He, Joeri Van~Mierlo, and Maitane Berecibar.
\newblock \capitalisewords{A Strategic Pathway from Cell to Pack-Level Battery Lifetime Model Development}.
\newblock {\em Applied Sciences}, 12(9):4781, 2022.

\bibitem{erdogan2022multi}
Nuh Erdogan, Sadik Kucuksari, and Jimmy Murphy.
\newblock \capitalisewords{A multi-objective optimization model for EVSE deployment at workplaces with smart charging strategies and scheduling policies}.
\newblock {\em Energy}, 254:124161, 2022.

\bibitem{ghosh2024isolating}
Sanchita Ghosh, Syed Ahsan~Raza Naqvi, Sai~Pushpak Nandanoori, and Soumya Kundu.
\newblock \capitalisewords{Isolating Signatures of Cyberattacks under Stressed Grid Conditions}.
\newblock {\em arXiv preprint arXiv:2408.02011}, 2024.

\bibitem{li2013reduced}
Xueyan Li, Meng Xiao, and Song-Yul Choe.
\newblock \capitalisewords{Reduced order model (ROM) of a pouch type lithium polymer battery based on electrochemical thermal principles for real time applications}.
\newblock {\em Electrochimica Acta}, 97:66--78, 2013.

\bibitem{li2020optimized}
Yunjian Li, Kuining Li, Yi~Xie, Jiangyan Liu, Chunyun Fu, and Bin Liu.
\newblock \capitalisewords{Optimized charging of lithium-ion battery for electric vehicles: Adaptive multistage constant current--constant voltage charging strategy}.
\newblock {\em Renewable energy}, 146:2688--2699, 2020.

\bibitem{sulzer2021python}
Valentin Sulzer, Scott~G Marquis, Robert Timms, Martin Robinson, and S~Jon Chapman.
\newblock \capitalisewords{Python battery mathematical modelling (PyBaMM)}.
\newblock {\em Journal of Open Research Software}, 9(1), 2021.

\bibitem{tranter2022liionpack}
Thomas Tranter, Robert Timms, Valentin Sulzer, Ferran Planella, Gavin Wiggins, Suryanarayana Karra, Priyanshu Agarwal, Saransh Chopra, Srikanth Allu, Paul Shearing, et~al.
\newblock \capitalisewords{liionpack: A Python package for simulating packs of batteries with PyBaMM}.
\newblock {\em Journal of Open Source Software}, 7(70), 2022.

\bibitem{jin2024assessing}
Lingkang Jin, Milad Kazemi, Gabriele Comodi, and Christina Papadimitriou.
\newblock \capitalisewords{Assessing battery degradation as a key performance indicator for multi-objective optimization of multi-carrier energy systems}.
\newblock {\em Applied Energy}, 361:122925, 2024.

\bibitem{chen2020development}
Chang-Hui Chen, Ferran~Brosa Planella, Kieran O’regan, Dominika Gastol, W~Dhammika Widanage, and Emma Kendrick.
\newblock \capitalisewords{Development of experimental techniques for parameterization of multi-scale lithium-ion battery models}.
\newblock {\em Journal of The Electrochemical Society}, 167(8):080534, 2020.

\bibitem{atlung1979dynamic}
Sven Atlung, Keld West, and Torben Jacobsen.
\newblock \capitalisewords{Dynamic aspects of solid solution cathodes for electrochemical power sources}.
\newblock {\em Journal of The Electrochemical Society}, 126(8):1311, 1979.

\bibitem{Ding}
Steven~X Ding.
\newblock {\em \capitalisewords{Model-based fault diagnosis techniques: design schemes, algorithms, and tools}}.
\newblock Springer Science \& Business Media, 2008.

\bibitem{khalid2021investigation}
Asadullah Khalid, Mohammad Khan, Alexander Stevenson, Shanzeh Batool, and Arif Sarwat.
\newblock \capitalisewords{Investigation of cell voltage buffer manipulation attack in a battery management system using unsupervised learning technique}.
\newblock In {\em 2021 IEEE Design Methodologies Conference (DMC)}, pages 1--6. IEEE, 2021.

\end{thebibliography}

\end{document}